\documentclass[12pt]{book}
\usepackage{plenum}
\usepackage{epsfig}
\usepackage{amssymb}
\begin{document}
\chapter{TOWARDS MATRIX MODELS OF IIB SUPERSTRINGS}

\author{P. Olesen} 

\affiliation{The Niels Bohr Institute\\
Blegdamsvej 17\\DK-2100 Copenhagen \O\ \\Denmark}

\vskip1cm

{\sl Invited talk at the ``Nato Advanced Research Workshop on Theoretical
Physics'', June 14-20, 1997, Zakopane, Poland}\\

\section{INTRODUCTION}

Recently there has been a lot of papers on matrix models and superstrings,
induced by the work of Banks, Fischler, Shenker, and Susskind (hep-th/
9706168). I refer to Makeenko's talk at this meeting for a general review
of this subject.  

Most of the work\footnote{The paper in reference 1 has been published in 
Nuclear Physics B. Unfortunately, the editors of that journal used an
early draft of the manuscript, which contained several typhos. For this reason
I cannot recommend the published version, but refer to the version in the
Archives.}  
 reported in this talk has been done together with 
Fayyazuddin, Makeenko, Smith, and Zarembo\refnote{1}. As explained in
Makeenko's (virtual) talk at this meeting, we started from the work by
Ishibashi, Kawai, Kitazawa, and Tsuchiya\refnote{2}, who proposed that type 
IIB superstrings in 10 dimensions are described by the reduced action,
\begin{equation}
S_{\rm IKKT}=\alpha \left(-\frac{1}{4}{\rm Tr}~[A_\mu,A_\nu]^2-\frac{1}{2}
{\rm Tr}~(\bar{\psi}\Gamma^\mu [A_\mu,\psi])\right),
\end{equation}
where $A_\mu$ and $\psi_\alpha$ are $n\times n$ matrices. A sum over $n$ is
implied, with weight $\exp (-\beta n)$. Later the sum over $n$ has been 
replaced by a double scaling limit\refnote{3}. 

In our paper\refnote{1} we discussed various problems associated with eq. (1),
and we proposed a different model with action
\begin{equation}
S_{\rm NBI}=-\frac{\alpha}{4}{\rm Tr}\left(Y^{-1}[A_\mu,A_\nu]^2\right)+
V(Y)-\frac{\alpha}{2}{\rm Tr}~(\bar{\psi}\Gamma^\mu[A_\mu,\psi]),
\end{equation}
where the potential is given by
\begin{equation}
V(Y)=\beta~{\rm Tr}Y+\gamma~{\rm Tr}\ln Y.
\end{equation}
The partition function is thus given by
\begin{equation}
Z=\int dA_\mu d\psi dY~e^{-S_{\rm NBI}}.
\end{equation}
We select the constant $\gamma$ in such a way that the result of the
$Y$-integration is as close to the superstring as is possible. This turns out
to mean
\begin{equation}
\gamma=n-\frac{1}{2},
\end{equation}
as we shall see later.

Physically the model  $S_{\rm NBI}$ is motivated by a GUT scenario: Suppose one
has a field theory valid down to the GUT scale. Then, in our model the group 
is SU($n$), with $n$ large. As we shall see, this type of GUT model then
leads to superstrings if $n\rightarrow\infty$. For $n$ finite, supersymmetry
is broken, as is expected for energies below the GUT energy.
Thus, superstrings can emerge
from a GUT type of model. Of course, the model with action $S_{\rm NBI}$ is not
a realistic GUT model.  

Under the SUSY transformations
\begin{equation}
\delta\psi=\frac{i}{4}\{Y^{-1},[A_\mu,A_\nu]\}\Gamma^{\mu\nu}\epsilon,~~
\delta A_\mu=i\bar{\epsilon}\Gamma_\mu\psi,
\end{equation}
the action transforms like
\begin{equation}
\delta S_{\rm NBI}\propto \epsilon^{\mu\alpha\beta\lambda_1...\lambda_7}~
{\rm Tr}\left(\psi_m(\Gamma^0\Gamma^{11}\Gamma^{\lambda_1}...
\Gamma^{\lambda_7})_{mp}\epsilon_p\{[A_\alpha,A_\beta],[A_\mu,Y^{-1}]\}\right).
\end{equation}
It can be shown that
\begin{equation}
\delta S_{\rm NBI}\rightarrow 0~~{\rm for}~~n\rightarrow\infty,
\end{equation}
so the action is supersymmetric in the limit $n\rightarrow\infty$, but for 
finite $n$ the symmetry is broken.

\section{THE $Y$-INTEGRATION}

The integration over $Y$ can be done exactly. Consider
\begin{equation}
{\cal F}(z)=\int dY~\exp \left(-\frac{\alpha}{4}{\rm Tr}(Y^{-1}z^2)-\beta~
{\rm Tr}Y-\gamma~{\rm Tr}\ln Y\right),~~z^2\equiv -[A_\mu,A_\nu]^2.
\end{equation}
The ``angular'' integration is of the Itzykson-Zuber type, so we get
\begin{equation}
{\cal F}(z)={\rm const.}\prod_{i=1}^{i=n}\int dy_i\frac{\Delta^2(y)}
{\Delta(1/y)\Delta(z^2)}~e^{-\alpha\sum_iz_i^2/4y_i-\beta\sum_iy_i
-\gamma\sum_i\ln y_i}.
\end{equation}
Here the $z_i$'s and $y_i$'s are the eigenvalues, and
\begin{equation}
\Delta(x)=\prod_{i>j}(x_i-x_j)=\det_{ki}x_i^{k-1}
\end{equation}
is the Vandermonde determinant. We only integrate over the positive
eigenvalues of $Y$. Thus we get
\begin{equation}
\Delta(z^2){\cal F}(z)={\rm const.}\int_0^\infty \prod_i dy_i ~y_i^{n-1}
\prod_{i>j}(y_i-y_j)~e^{-\alpha\sum_iz_i^2/4y_i-\beta\sum_iy_i
-\gamma\sum_i\ln y_i}.
\end{equation}
This can be rewritten as a determinant
\begin{eqnarray}
\Delta(z^2){\cal F}(z)&=&{\rm const.}~\det_{ki}\int \frac{dy}{\sqrt{y}}
y^{k-1}e^{-\alpha z_i^2/4y-\beta y}\nonumber \\
&=&{\rm const.}~\det_{ki}\left[(-1)^{k-1}\frac{\partial^{k-1}}{\partial 
\beta^{k-1}}
\left(\sqrt{\frac{\pi}{\beta}}~e^{-\sqrt{\alpha\beta}z_i}\right)\right].
\end{eqnarray}
Here $z_i$ is by definition the positive square root of $z_i^2$.
This determinant can be evaluated using basic properties of determinants,
and the result is\refnote{1}
\begin{equation}
\Delta(z^2){\cal F}(z)={\rm const.}~\Delta(z)~e^{-\sqrt{\alpha\beta}~\sum_i
z_i}.
\end{equation}
This result is exact, and hence it is valid for any $n$.

The sum over eigenvalues in the exponent has the following interpretation,
\begin{equation}
\sum_iz_i=\sum_i\sqrt{z_i^2}=\frac{1}{4i\sqrt{\pi}}\int_{-\infty}^{(0+)}
\frac{dt}{t^{3/2}}\sum_ie^{tz_i^2},
\end{equation}
where we used an integral representation of the square root. Thus,
\begin{equation}
\sum_iz_i=\frac{1}{4i\sqrt{\pi}}\int_{-\infty}^{(0+)}~\frac{dt}{t^{3/2}}~
{\rm Tr}~\exp (-t[A_\mu,A_\nu]^2)={\rm Tr}~\sqrt{-[A_\mu,A_\nu]^2},
\end{equation}
valid in Euclidean space.

The partition function therefore becomes
\begin{eqnarray}
Z&=&\int dA_\mu d\psi dY~\exp \left(\frac{\alpha}{4}{\rm Tr}(Y^{-1}[A_\mu,
A_\nu]^2)-\beta~{\rm Tr}Y-\gamma~{\rm Tr}\ln Y\right)\nonumber \\
&=&{\rm const.}\int\frac{ dA_\mu d\psi}
{\prod_{i>j}(z_i+z_j)}
\exp \left(-\sqrt{\alpha\beta}~{\rm Tr}\sqrt{-[A_\mu,A_\nu]^2}-
\frac{\alpha}{2}{\rm Tr}~(\bar{\psi}\Gamma^\mu[A_\mu,\psi])\right).\nonumber \\
\end{eqnarray}
This is the exact result of the $Y$-integration. In order to get the square 
root it is important to use the value of $\gamma$ given in eq. (5).

This result can be expressed in an alternative form, at the cost of introducing
an auxillary Hermitean field $M$. We use the identity
\begin{equation}
\frac{1}{\prod_{i<j}(z_i+z_j)}={\rm const.}\sqrt{\det z}\int dM
e^{-{\rm Tr}~z M^2},
\end{equation}
to obtain
\begin{eqnarray}
Z&=&{\rm const.}\int dA_\mu d\psi dM
\left(\det\sqrt{-[A_\mu,A_\nu]^2}\right)^{1/2}\nonumber \\
&&\times \exp \left(-{\rm Tr}\left((\sqrt{\alpha\beta}+M^2)
\sqrt{-[A_\mu,A_\nu]^2}\right)-
\frac{\alpha}{2}{\rm Tr}~(\bar{\psi}\Gamma^\mu[A_\mu,\psi])\right).
\end{eqnarray}
The field $M$ is essentially trivial, with a ``classical equation of motion''
$M=0$.

\section{ON THE WEYL REPRESENTATION AND THE APPROACH OF 
THE COMMUTATOR TO THE POISSON BRACKET}

The square root occuring in the result above is somewhat reminisent of the 
Nambu-Goto square root. If we could replace the commutator in the square root
in eq. (17) by the corresponding Poisson bracket, we would have a partition
function which is very similar to the one for the superstring. 

This problem has been discussed by Hoppe\refnote{4}, and in different 
settings by a number of other authors\refnote{5}$^,$\refnote{6}$^,$
\refnote{7}. It turns out that making some assumptions, one has the
limit
\begin{equation}
``~[A,B]\rightarrow i\{A,B\}_{\rm PB}~``,~~{\rm for}~~n\rightarrow\infty.
\end{equation}
Here $\{,\}_{\rm PB}$ denotes the usual Poisson bracket.

We refer to the literature for a detailed discussion. Here we shall follow
Bars\refnote{7}, and consider a torus (although this restriction is probably 
not important\refnote{8}). The case of a sphere was discussed in ref. 5. 
A Hermitean matrix can be expanded in a Weyl basis,
\begin{equation}
(A_\mu)^i_j=C_1\sum_{\bf k}a_\mu^{\bf k}(l_{\bf k})^i_j,~~{\rm with}~~
{\bf k}=(k_1,k_2).
\end{equation}
The matrix $l_{\bf k}$ can be expressed in terms of the $n\times n$ ($n=$odd)
Weyl matrices $h$ and $g$, which satisfy
\begin{equation}
h^n=g^n=1~~{\rm and}~~gh=\omega hg,~~{\rm with}~~\omega=e^{4\pi i/n}.
\end{equation}
The explicit form of these matrices are
\begin{eqnarray}
h&=&{\rm diag}(1,\omega,\omega^2,...,\omega^{n-1})\nonumber \\
g^i_j&=&\delta^i_{j+1},~i=1,2,...,n-1,~g^n_j=0~{\rm except ~for}~g^n_1=1.
\end{eqnarray}
The SU($n$)-generators $l_{\bf k}$ are then constructed as
\begin{equation}
l_{\bf k}=\frac{n}{4\pi}\omega^{k_1k_2/2}h^{k_1}g^{k_2},
\end{equation}
since the powers of $h$ and $g$ are linearly independent for $k_1,k_2=1,2,...
,n-1$, are unitary, close under multiplication, and are traceless. Using 
that
\begin{equation}
{\rm Tr}~h^{k_1}g^{k_2}=n\delta_{k_1,0}\delta_{k_2,0},
\end{equation}
we easily see that
\begin{equation}
{\rm Tr}~l_{\bf k}l_{\bf p}=\frac{n^3}{(4\pi)^2}\delta_{\bf k+p,0}.
\end{equation}
Thus the expansion coefficients in eq. (21) are given by
\begin{equation}
a_\mu^{\bf-p}=(a_\mu^{\bf p})^*=\frac{(4\pi)^2}{n^3C_1}{\rm Tr}~(l_{\bf 
p}A_\mu).
\end{equation}
Also, from the relation
\begin{equation}
g^{k_1}h^{k_2}=\omega^{k_1k_2}h^{k_2}g^{k_1}
\end{equation}
we get
\begin{equation}
[l_{\bf p},l_{\bf k}]=i\frac{n}{2\pi}~\sin\left(\frac{2\pi}{n}{\bf p\times k}
\right)l_{\bf p+k},
\end{equation}
where
\begin{equation}
{\bf p\times k}=p_1k_2-k_1p_2.
\end{equation}

Using the expansion (21) we get
\begin{equation}
[A_\mu,A_\nu]_i^j=C_1^2\sum_{\bf p,q}a_\mu^{\bf p}a_\nu^{\bf q}\frac{n}{2\pi}
\sin\left(\frac{2\pi}{n}{\bf p\times q}\right)(l_{\bf p+q})_i^j.
\end{equation}
This can be compared with the similar expression for the string variables
$X_\mu(\sigma,\tau)$, where we have the expansion
\begin{equation}
X_\mu(\sigma,\tau)=C_2\sum_{\bf m}a_\mu^{\bf m}\exp(i\sigma m_1+i\tau m_2),
\end{equation}
leading to
\begin{equation}
\{X_\mu,X_\nu\}_{\rm PB}=C_2^2\sum_{\bf p,q}a_\mu^{\bf p}a_\nu^{\bf q}~
({\bf p\times q})\exp(i({\bf p+q})\sigma).
\end{equation}

Now it is clear that {\sl if}
\begin{equation}
\lim_{n\rightarrow\infty}\sum_{\rm modes}...=\sum_{\rm modes}\lim_{n\rightarrow
\infty}...,
\end{equation}
then we have by use of (26)
\begin{equation}
{\rm Tr}~[A_\mu,A_\nu]^2\rightarrow {\rm const.} \int d\sigma d\tau
\{X_\mu,X_\nu\}^2~~{\rm for}~~n\rightarrow\infty.
\end{equation}
It is obvious that the commutativity (34) is only valid if the infinite modes
are unimportant. This is, however, not true e.g. for the bosonic string.
If we fix the end points of this string at some distance, then there is
a critical distance (essentially the inverse tachyon mass) at which the 
string oscillates so wildly that this behaviour can only be reproduced with 
an infinite number of modes. Below this distance the ``string'' becomes
a branched polymer, and hence is no longer a string.

For superstrings this problem does not arize, and hence there is at least
no obvious reason why the limits cannot be intechanged as in eq. (35).
In the following we assume that eq. (35) is correct for type IIB superstrings.

Since we are interested in the square root of the squared commutator, the
result (35) is not enough. Using eq. (26) and repeated applications of the 
formula
\begin{equation}
l_{\bf m}l_{\bf r}=(n/4\pi)\exp(2\pi i({\bf m\times r})/n))l_{\bf m+r}
\end{equation}
one can easily derive
\begin{equation}
{\rm Tr}~l_{\bf m_1}l_{\bf m_2}...l_{\bf m_s}\rightarrow n(n/4\pi)^s~
\delta_{\bf m_1+m_2+...+m_s,0},~{\rm for}~n\rightarrow\infty
\end{equation}
to leading order in $n$. Using this in eq. (16) we obtain
\begin{eqnarray}
{\rm Tr}~\sqrt{-[A_\mu,A_\nu]^2}&=&
\frac{1}{4i\sqrt{\pi}}\int_{-\infty}^{(0+)}~\frac{dt}{t^{3/2}}{\rm Tr}~
\exp (-t[A_\mu,A_\nu]^2)\nonumber \\
&\rightarrow&\frac{1}{4i\sqrt{\pi}}\int_{-\infty}^{(0+)}~\frac{dt}{t^{3/2}}
\int d\sigma d\tau \exp (t~{\rm const.}\{X_\mu,X_\nu\}_{\rm PB}^2)\nonumber \\ 
&=&{\rm const.}\int d\sigma d\tau
\sqrt{\{X_\mu,X_\nu\}_{\rm PB}^2}~~{\rm for}~n\rightarrow\infty.
\end{eqnarray}
Thus we see that in the leading order the Nambu-Goto square root arises
as the limit of the square root of the corresponding commutator. However, it
should be remembered what was said before about strings with tachyons. They
do not allow the interchange of limits as in (34), and hence the result
(38) is not valid in that case\footnote{For such strings where the end points
are actually separated by a {\sl large} distance, the interchange of limits in
(34) is probably allowed. Thus at large distances the string picture is most 
likely right for the NBI matrix model even without supersymmetry. At shorter
distances near the critical one, this picture breaks down, and the sine
function in (31) cannot be approximated by its first term in a power series
expansion.}.

\section{TYPE IIB SUPERSTRING FROM THE NBI MATRIX MODEL}

We can now summarize our results in the following rather long formula,
\begin{eqnarray}
Z&=&\int dA_\mu d\psi dY~\exp \left(\frac{\alpha}{4}{\rm Tr}(Y^{-1}
[A_\mu,A_\nu]^2)-\beta~{\rm Tr}Y-(n-\frac{1}{2})~{\rm Tr}\ln Y\right)
\nonumber \\
&=&{\rm const.}\int dA_\mu d\psi dM
\left(\det\sqrt{-[A_\mu,A_\nu]^2}\right)^{1/2}\nonumber \\
&&\times \exp \left(-{\rm Tr}\left((\sqrt{\alpha\beta}+M^2)
\sqrt{-[A_\mu,A_\nu]^2}\right)-
\frac{\alpha}{2}{\rm Tr}~(\bar{\psi}\Gamma^\mu[A_\mu,\psi])\right)\nonumber \\
&\rightarrow&{\rm const.}\int dX_\mu d\psi dM
\left(\det\sqrt{\{X_\mu,X_\nu\}^2}\right)^{1/2}\nonumber \\
&&\times \exp \left(-\int d\sigma d\tau\left((\sqrt{\alpha\beta}+M^2)
\sqrt{\{X_\mu,X_\nu\}^2}-
\frac{i\alpha}{2}(\bar{\psi}\Gamma^\mu\{X_\mu,\psi\})\right)\right),
\end{eqnarray}
where the last expression is valid for $n\rightarrow\infty$. In this expression
the fields $\psi$ and $M$ have expansions similar to eqs. (21) and (32), and
some normalization constants have been absorbed in $\alpha$ and $\beta$ in the
last formula above.

The functional integration over $M$ in eq. (39) is just Gaussian and can of 
course easily be performed,
\begin{eqnarray}
z&\rightarrow&{\rm const.}\int dX_\mu d\psi 
\left(\frac{\det\sqrt{\{X_\mu,X_\nu\}^2}}
{{\cal D}et\sqrt{\{X_\mu,X_\nu\}^2}}\right)^{1/2}\nonumber \\
&&\times \exp \left(-\int d\sigma d\tau \left(\sqrt{\alpha\beta}
\sqrt{\{X_\mu,X_\nu\}^2}-
\frac{i\alpha}{2}(\bar{\psi}\Gamma^\mu\{X_\mu,\psi\})\right)\right).
\end{eqnarray}
The two determinants in this expression arises from different types of Gaussian
integrations, the ``det'' beeing defined through eq. (18) and the subsequent
limit $n\rightarrow\infty$, whereas the ``${\cal D}et$'' determinant comes from
the continuum integral over $M$. Naively one would tend to identify these 
two determinants, so that the fraction containing them is just one,
\begin{equation}
\frac{\det\sqrt{\{X_\mu,X_\nu\}^2}}
{{\cal D}et\sqrt{\{X_\mu,X_\nu\}^2}}\rightarrow 1.
\end{equation}
If so, the NBI action gives exactly the Nambu-Goto version of the 
Green-Schwarz type IIB superstring.

However, Zarembo\refnote{9} has pointed out to me that the situation can be
more complicated. For example, from eq. (18) one sees that the ``det''
determinant ($=\det z$ for $n\rightarrow\infty$) is subdominant relative
to the factor $\prod (z_i+z_k)$ occurring in eq. (18), and hence can be
ignored in the limit $n\rightarrow\infty$. In this case, one has instead of
(41)
\begin{equation}
\frac{\det\sqrt{\{X_\mu,X_\nu\}^2}}
{{\cal D}et\sqrt{\{X_\mu,X_\nu\}^2}}\rightarrow \frac{1}
{{\cal D}et\sqrt{\{X_\mu,X_\nu\}^2}}.
\end{equation}
If so, the ${\cal D}et$ determinant survives in the measure. 
However, even if this is so, this factor is rather harmless: correlation 
functions are invariant, since the measure is multiplied by a constant factor
under reparametrizations. This factor does not depend on the fields and
cancels in the correlation functions\refnote{9}.

Perhaps the right answer depends on how exactly the continuum limit is
constructed, because in order to interpret ${\cal D}et$ a regulator is
needed.

It should also be mentioned that Chekhov and Zarembo\refnote{10} have
discussed models somewhat different from the NBI model, and have also
discussed the measure in more details.

\section{A SADDLE POINT AND THE VIRTUAL EULER NUMBER}

We shall now study the saddle point of the NBI action. By variation of
the $A_\mu-$fields we obtain the classical equation of motion
\begin{equation}
[A_\mu,\{Y^{-1},[A_\mu,A_\nu]\}]=0.
\end{equation}
This equation was studied by Kristjansen and me\refnote{11}. The solution is
\begin{equation}
[A_\mu,A_\nu]_j^i=im_{\mu\nu}~Y_j^i,
\end{equation}
where $m_{\mu\nu}$ is a matrix with repect to the space indices. In the saddle
point the action has the value
\begin{equation}
S_{\rm NBI}^{\rm saddle}=(\beta+m_{\mu\nu}^2\alpha/4)~{\rm Tr}Y
+(n-1/2)~{\rm Tr}\ln Y.
\end{equation}
In order to have a non-trivial $n\rightarrow\infty$ limit, it is necessary that
$\alpha$ and $\beta$ are of order $n$. It should be stressed that this does
{\sl not} imply the usual classical limit in string theory, as explained
in details in ref. 11. 

In addition to the terms exhibited above, there are of course subdominant terms
arising from the expansion of $A_\mu$ around the classical solution. These
terms are ignored in the following. Therefore, at the $A_\mu$-saddle point we 
have the integral\refnote{11} ($\alpha/n$ and $\beta/n$ are both of order one)
\begin{equation}
Z^{\rm saddle}=\int dY \exp [-n\left\{(\beta+m_{\mu\nu}^2\alpha/4)/n~{\rm Tr}Y
+t~{\rm Tr}\ln Y\right\}],~{\rm with}~~t=1-1/2n.
\end{equation}
This functional integral is of the Penner type\refnote{12}. For the value of 
the parameter $t$ needed in the saddle point, the $Y$-integral actually
diverges. However, by analytic continuation \'a la the gamma function for
negative argument one can start by defining the integral for negative $t$,
and then ultimately continue back to positive $t$. It the turns out that $t=1$
is a critical point, and in the vicinity of this point one can define a
double scaling limit with the ``cosmological constant''
\begin{equation}
\mu=(1-t)n={\rm fixed}.
\end{equation}
We see that with the value of $t$ given by eq. (46), $\mu=$1/2=fixed quite
automatically! Thus, we do not need to make any special assumptions in
order to have this double scaling limit in the NBI model.

What is the meaning of the Penner model in the double scaling limit? An
asymptotic expansion in $\mu$ can be made. Consider the ``free energy'' $F$,
$Z^{\rm saddle}\equiv e^F$, then
\begin{equation}
F(\mu)=F_0(\mu)+F_1(\mu)+\sum_{g=2}^{\infty}\chi_g\mu^{2-2g},~~\mu=1/2.
\end{equation}
Here $\chi_g$ is the ``virtual Euler number'' for moduli space of Riemann
surfaces with genus $g$, which is well known to be relevant for strings. One 
has\refnote{12}
\begin{equation}
\chi_g=\frac{B_{2g}}{2g(2g-2)},
\end{equation}
where $B_{2g}$ are the Bernoulli numbers. These have positive sign, and blow up
factorially, so the sum defining $F$ is not Borel summable. This is also
well known to be the case for genus expansion of string theories.

The physical interpretation of this result is that the field $Y$ captures the
Euler characteristic of moduli space of Riemann surfaces. Therefore it is
quite likely that the NBI model encodes non-perturbative information on
Riemann surfaces generated by moduli space. It should be remembered that
in string theories one usually sum the functional integral over $g$, however 
here this seems to be already included. It must be admitted that the virtual 
Euler number represents very global properties of moduli space, and
certainly more details are needed before one can claim a good understanding
of the non-perturbative nature of this model.

Recently Soloviev\refnote{13} has commented on ``a curious relation'' between
Siegel's model\refnote{14} of random lattice strings and the above saddle point
approximation to the NBI model. This comes about if one starts from Siegel's
T-self-dual matrix model
\begin{equation}
S={\rm Tr}\left(\frac{1}{2}\Phi^2+n~\ln(1-g\Phi)\right),
\end{equation}
where $\Phi$ is a Hermitean $n\times n$ matrix and $g$ is a constant. It was
then pointed out by Soloviev\refnote{13} that if one makes the substitution
\begin{equation}
gY=1-g\Phi,
\end{equation}
and perform the limit $n\rightarrow\infty,~ g\rightarrow 0,~ gn=$fixed,
then one obtains
\begin{equation}
S\rightarrow n~{\rm Tr ~(const.}Y+n\ln Y)+{\rm irrelevant~ const.}.
\end{equation}
This is, however, precisely the saddle point expression (46) for the NBI
model. This sadle point is therefore a weak string coupling limit of the
Siegel matrix model\refnote{13}. For arbitrary coupling there is,
however, an additional $Y^2$-term in the Siegel action, and hence it was
suggested that perhaps the potential (3) should have an additional
Tr$Y^2$ term\refnote{13}. Of course, a similar statement can be made about the
NBI model, where there are various corrections to the saddle point
expansion.

%\refnote{1}

\end{document}